\documentclass[journal]{new-aiaa}

\usepackage{amsthm}
\usepackage[utf8]{inputenc}
\usepackage{textcomp}
\usepackage[dvipsnames]{xcolor}
\usepackage{graphicx}
\usepackage{amsmath}

\usepackage[version=4]{mhchem}
\usepackage{siunitx}
\usepackage{longtable,tabularx}
\usepackage{algorithm}
\usepackage{algpseudocode}
\newcommand{\T}{^\mathrm{T}}

\DeclareMathOperator*{\argmax}{arg\,max}

\newcommand{\vn}{{\mathbf n}}

\newcommand{\vr}{{\mathbf r}}

\newcommand{\vu}{{\mathbf u}}

\newcommand{\vx}{{\mathbf x}}

\newcommand{\vI}{{\mathbf I}}

\newcommand{\vT}{{\mathbf T}}

\newcommand{\vZ}{{\mathbf Z}}

\newcommand{\Eb}{\mathbb{E}}

\newcommand{\Qb}{\mathbb{Q}}

\newcommand{\rd}{\mathrm{d}}

\newtheorem{lemma}{Lemma}
\newtheorem{definition}{Assumption}

\title{Deep $\mathcal{L}^1$ Stochastic Optimal Control Policies for Planetary Soft-landing}

\author{Marcus A. Pereira \footnote{Ph.D. student in Robotics at Georgia Tech, email address: mpereira30@gatech.edu}} 
\affil{The Institute for Robotics and Intelligent Machines, Georgia Institute of Technology, Atlanta, GA 30332}
\author{ Camilo A. Duarte\footnote{Master's student in the School of Aerospace Engineering at Georgia Tech, email address: candresdu@gmail.com}}
\affil{School of Aerospace Engineering, Georgia Institute of Technology, Atlanta, GA 30332}
\author{Ioannis Exarchos\footnote{Work done during a postdoctoral fellowship at Stanford University, %Currently, Data \& Applied Scientist at Microsoft, 
email address: exarchos@gatech.edu
}}
\affil{Microsoft}
\author{Evangelos A. Theodorou\footnote{Associate Professor at the Daniel Guggenheim School of Aerospace Engineering, email address: evangelos.theodorou@gatech.edu}}
\affil{School of Aerospace Engineering, Georgia Institute of Technology, Atlanta, GA 30332}

\begin{document}

\maketitle

\begin{abstract}
In this paper, we introduce a novel deep learning based solution to the Powered-Descent Guidance problem, grounded in principles of nonlinear Stochastic Optimal Control and Feynman-Kac theory. Our algorithm solves the PDG problem by framing it as an $\mathcal{L}^1$ SOC problem for minimum fuel consumption. Additionally, it can handle practically useful control constraints, nonlinear dynamics and enforces state constraints as soft-constraints. This is achieved by building off of recent work on deep Forward-Backward Stochastic Differential Equations and differentiable non-convex optimization neural-network layers based on stochastic search. In contrast to previous approaches, our algorithm does not require convexification of the constraints or linearization of the dynamics and is empirically shown to be robust to stochastic disturbances and the initial position of the spacecraft. After training offline, our controller can be activated once the spacecraft is within a pre-specified radius of the landing zone and at a pre-specified altitude i.e., the base of an inverted cone with the tip at the landing zone. We demonstrate empirically that our controller can successfully and safely land all trajectories initialized at the base of this cone while minimizing fuel consumption.          
\end{abstract}

% \section*{Nomenclature}
% {\renewcommand\arraystretch{1.0}
% \noindent\begin{longtable*}{@{}l @{\quad=\quad} l@{}}
% $t_f$ & Simulation time or time horizon in seconds \\
% $t$ & Time in seconds, with $t \in [0,\,t_f]$\\
% $\vx(t)$  & State vector\\
% $\vu(t)$ & Controls vector \\
% $V(\vx,\,t)$ & Value function starting from state $\vx$ and time $t$ \\
% $V_\vx(\vx,\,t)$  & Gradient of the value function\\
% $V_{\vx\vx}(\vx,\,t)$ & Hessian of the value function\\
% $\mathcal{H}$ & Hamiltonian, is a function of $t,\,\vx(t),\,\vu(t),\,V_\vx(\vx,\,t)$ and $V_{\vx\vx}(\vx,\,t)$
% \end{longtable*}}

\section{Introduction and Related Work}
The Powered-Descent Guidance (PDG) problem addresses the final stage of entry, descent, and landing sequence wherein a spacecraft uses its rocket engines to maneuver from some initial position to a soft-landing at a desired landing location. It can be framed as a finite time-horizon optimal control problem where the ultimate goal is to achieve a safe landing while minimizing the amount of fuel consumed during descent. The definition of a safe landing is provided in terms of state constraints (such as terminal velocity and position) derived from mission critical requirements. As a consequence, PDG is regarded as a control- and state-constrained optimization problem, with state constraints imposed by stringent mission requirements and control constraints imposed by the thrusting capabilities of the spacecraft. The PDG problem is commonly framed as an $\mathcal{L}^1$ optimal control problem \cite{exarchos2019optimal} wherein the $\mathcal{L}^1$-norm of the control is used instead of the standard quadratic control cost. This typically results in a \textit{max-min-max} thrust profile instead of continuous thrusting as prescribed by a quadratic control cost minimizing controller.
\par \textbf{Motivation for using the $\mathcal{L}^1$-norm:} The cost of fuel in space is exponentially larger than any other terrestrial application. Thus, minimizing fuel consumption becomes a critical component in the design of cost functions for the PDG optimal control problem. The fallacy of the assumption that \textit{quadratic costs minimize fuel-consumption} is proved in \cite{imross2004}. In this work, the author demonstrates how the choice of the norm of the thrust in the cost function is dependent on the type of rocket and which norms actually measure fuel consumption. It is shown that the well-known quadratic cost (or $\mathcal{L}^2$-norm) does not measure (and therefore does not minimize) fuel consumption and that a control policy optimal for quadratic costs will be sub-optimal with respect to other control costs that do measure fuel-consumption. 
Additionally, as mentioned in \cite{imross2004}, continuous thrusting controllers (obtained from quadratic costs), can cause undesirable effects (such as increasing the microgravity environment) on precision pointing payloads. For such payloads, \textit{bang-off-bang} controllers are preferable so that science can happen during the \textit{off} periods. Thus, the $\mathcal{L}^1$-norm is the \textit{de facto} choice for designing optimal controllers for space applications. 

\par \textbf{Related work:} The PDG optimal control problem is a non-convex optimization problem. One approach is to convexify the original problem and prove that the convexification is lossless \cite{dueri2017customized}. However, proving this is not trivial and requires assumptions leading to ignoring certain constraints (such as the descent glide-slope) that help simplify the analysis. These also require linearizing the dynamics and deriving subsequent error bounds. However, the advantages are that it allows using \textit{off-the-shelf} convex programming solvers and guarantees unique solutions. Another approach is to use sequential convex programming to iteratively convexify the original problem \cite{reynolds2020real}. Moreover, these approaches consider deterministic dynamics (i.e., cannot handle stochastic disturbances or unmodeled phenomena) and solve the problem for a specific initial condition. To handle stochasticity or arbitrary initial conditions, the solutions have to be recomputed \textit{on-the-fly}. The authors in \cite{ridderhof2019minimum} consider a stochastic version of the PDG problem, however, they do not consider stochasticity in the dynamics of the mass of the spacecraft. As will be seen in our problem formulation, the stochasticity entering the mass dynamics are negatively correlated to that entering the acceleration dynamics. Additionally, to handle the non-convex thrust-bounds constraint, they impose a control structure allowing Gaussian controls and then constrain only the mean to satisfy conservative thrust bounds. This makes the problem deterministic and the same lossless convexification solution as in \cite{dueri2017customized} can be used. However, the conservative bounds lead to increased fuel consumption for which they propose solving an additional covariance steering problem. This solution relies on linear dynamics and does not work when there is stochasticity in the mass dynamics and the state vector contains the spacecraft's mass thus yielding a nonlinear dynamical model.  Another approach \cite{exarchos2019optimal} based on the same stochastic optimal control theory as ours, presents a solution for the one-dimensional stochastic PDG problem. However, the closed-form optimal control expression presented in this work does not hold for the general three-dimensional constrained PDG problem as well as the proposed numerical algorithm is prone to compunding errors from least-squares approximations at every time step. Nevertheless, the results in terms of crash percentages demonstrate superior performance to deterministic controllers as well as the venerable Apollo powered descent guidance law and comparable performance in terms of fuel consumption. This motivates our work based on the same theory but delivers a general solution.
\par There are recent works in literature that use deep neural networks (DNNs) to solve the deterministic soft landing problem. In \cite{sanchez2018real}, the authors employ an imitation learning-like procedure wherein Pontryagin's Maximum Principle (PMP) is used to solve optimal control problems for soft-landing and generate training data. This data is then used to train DNNs via supervised learning. The authors claim that the learned policy can generalize to unseen areas of the state space. However, their approach considers a two-dimensional representation of a rocket and does not consider any state constraints. In \cite{you2020learningbased}, the authors solve the 2D PDG problem for a spacecraft with vectorized thrust by formulating a Hamiltonian through the use of PMP and derive the necessary conditions of optimality that lead to a Two-Point Boundary Value Problem. They use a DNN to approximate the initial conditions of the adjoint variables which are then used to forward propagate the adjoint variables in time. Our proposed solution using deep Forward-Backward Stochastic Differential Equations (FBSDEs) adopts a similar strategy to allow forward-propagation of the backward SDE (BSDE).
\par To the best of our knowledge, our work is the first to propose a deep learning based solution to the stochastic three dimensional constrained PDG problem. Our work is inspired by \cite{exarchos2019optimal} and builds off of recent work \cite{pereira2019learning, exarchos2020novas} that use DNNs to solve systems of FBSDEs. These so called deep FBSDE controllers are scalable solutions to solve high-dimensional parabolic partial differential equations such as the Hamilton-Jacobi-Bellman (HJB) PDE that one encounters in continuous-time stochastic optimal control problems. These do not suffer from compounding least-squares errors and do not require backpropagating SDEs. By treating the initial-value of the BSDE as a learnable parameter of the DNN, the BSDE can be forward propagated and the deviation from the given terminal-value can be used as a loss function to train the DNN. These controllers have been used to successfully solve high-dimensional problems in finance \cite{exarchos2020novas} and safety-critical control problems \cite{pereira2020safe}. Compared to the work thus far on deep FBSDEs and PDG literature, our main contributions are as follows: 
\begin{enumerate}
    \item Ability to solve the nonlinear $\mathcal{L}^1$ Stochastic Optimal Control PDG problem using deep FBSDEs without relying on convexification and convex solvers in an end-to-end differentiable manner.  
    \item Incorporated \textit{first-exit} time capability into the deep FBSDE framework for the PDG problem. 
    \item Can be trained to be invariant of the initial position of the spacecraft and handle stochastic disturbances. The trained network can be deployed as a feedback policy without having to recompute the optimal solution online. 
\end{enumerate}
With regards to computational burden, similar to \cite{sanchez2018real}, our approach is also based on training a policy network offline. The online computation comprises of a forward pass through a neural network and one-step parallel simulation of the dynamics. These computations can be performed entirely on a CPU (using vectorized operations) or a modest GPU.

\section{Problem Formulation}
 In this section, we present the dynamics of the spacecraft, the control and state constraints generally considered for soft-landing and how we handle them and finally the PDG stochastic optimal control problem for which we propose an algorithm and an empirical solution in subsequent sections.

\subsection{Spacecraft Dynamics and Constraints}
For our purposes, we make the following assumptions: (1) aerodynamic forces are neglected such that only gravity and thrust forces act on the vehicle, (2) the spacecraft is at a relatively low altitude (final stage of descent) such that a flat planet model can be assumed, and at a reasonable distance to the desired landing zone; (3) similar to \cite{dueri2017customized} we assume high bandwidth attitude control so that we can decouple translational and rotational dynamics and (4) we consider the initial velocity to be in the subsonic regime. Due to the assumption (3), we completely neglect rotational dynamics of the spacecraft in this formulation and assume that the attitude of the vehicle needed to produce the required thrust profile can be achieved instantaneously. Therefore, it is sufficient to define the dynamics of the vehicle by its translational dynamics which are as follows:\begin{align}
\begin{split}
    \label{eqn:dynamics}
    \dot{\textbf{r}}(t) &= \textbf{v}(t), \\ 
    \dot{\textbf{v}}(t) &= \frac{\textbf{T}(t)}{m(t)}  - \textbf{g} \\
    \dot{m}(t) &= - \alpha ||\textbf{T}(t)||
\end{split}
\end{align} where, at time $t$, $\textbf{r}(t) \in \mathbb{R}^3$ is the position of the spacecraft  with respect to a defined inertial frame, $\textbf{v}(t) \in \mathbb{R}^3$ is the velocity defined in the same frame and $m(t) \in \mathbb{R}^+$ is the spacecraft's total mass. $\textbf{T} \in \mathbb{R}^3$ is the thrust vector generated by the propulsion system, $\textbf{g} \in \mathbb{R}^3$ is the acceleration vector due to the gravitational force exerted by the planet on the spacecraft, and $\alpha \in \mathbb{R}^+$ governs the rate at which fuel is consumed with the resulting generated thrust. Hereon, thrust $\vT(t)$ and control $\vu(t)$ will be used interchangeably. 
\par In a stochastic setting, as described in \cite{exarchos2019optimal}, we assume that stochastic disturbances enter the acceleration channels due to unmodeled environmental disturbances and also because we can assume uncertainty in the exact thrust value exerted by the spacecraft due to limitations in the precision of our control. Moreover, these disturbances are negatively correlated with the noise that enters the mass-rate channel. Thus, we have the following stochastic dynamics, 
\begin{align}
\begin{split}
    \label{eqn:sde_dynamics}
    \rd \vr(t) &= \textbf{v}(t) \rd t, \\ 
    \rd \textbf{v}(t) &= \bigg[ \frac{\textbf{T}(t)}{m(t)}  - \textbf{g}\bigg]\rd t + \frac{\Gamma}{m(t)} \rd W(t), \\
    \rd {m}(t) &= - \alpha \bigg [ ||\textbf{T}(t)||\rd t + \mathbf{1}_{1 \times 3}\T\, \Gamma\, \rd W(t) \bigg ]
    %TODO: Fix Sigma here
\end{split}
\end{align}
where, $\rd W \in \mathbb{R}^3$ is a vector of mutually independent Brownian motions and $\Gamma \in \mathbb{R}^{3 \times 3}$ is a diagonal matrix of variances implying that the noise entering the three acceleration channels are uncorrelated. A column vector of ones ($\mathbf{1}_{1 \times 3}$) is used to combine the Brownian motions in the acceleration channels to obtain a Brownian motion that enters the mass-rate channel which is negatively correlated with those that enter the acceleration channels (due to the $-\alpha$ coefficient). We can rewrite the dynamics concisely as a stochastic differential equation as follows:\begin{equation}
    \rd \vx(t) = f(\vx(t),\,\vT(t)) \,\rd t + \Sigma(\vx(t)) \,\rd W(t),
\end{equation}
where, $\vx(t)\in \mathbb{R}^7$ is the state vector,  $f(\vx(t),\,\vT(t))$ is the \textit{drift} vector representing the deterministic component and $\Sigma(\vx(t)) \triangleq H(\vx(t)) \Gamma $ is the \textit{diffusion} matrix representing the stochastic component of the dynamics. The state ($\vx(t)$) is defined as, $\vx = [\vr(t)\T,\, \mathbf{v}(t)\T,\, m(t)]\T$ and $H(\vx)$ is a $7 \times 3$ matrix defined as follows,\begin{align*}
    H(\vx(t)) = 
    \begin{bmatrix}
    \mathbf{0}_{3 \times 3} & 
    \frac{1}{m(t)} \mathbf{I}_{3 \times 3} & 
    - \alpha \mathbf{1}_{3 \times 1}
    \end{bmatrix} ^T
\end{align*} We first begin with the control constraints that are generally considered in PDG problems. These are imposed by physical limitations on the spacecraft's propulsion system. In order for the propulsion system to operate reliably, the engines may not operate below a certain thrust level. We also know that, realistically, the thrusters are only capable of producing finite thrust. These are enforced by the following constraint,\begin{align}
\label{eqn:thrust_bounds}
    0 < \rho_1 \leq ||\mathbf{T}(t)|| \leq \rho_2
\end{align}
\noindent 
This constraint leads to a non-convex set of feasible thrust values due to the lower-bound. The conventional approach \cite{dueri2017customized} is to convexify the problem to handle the non-convex constraints and show that the convexification is losses. In this paper, we will work directly with the non-convex constraints.    
\par Additionally, a constraint on the direction in which thrust can be applied is also imposed. The so-called \textit{thrust-pointing} constraint is given by,
\begin{align}
\label{eqn:thrust_pointing}
    \hat{\mathbf{n}} \cdot \mathbf{T}(t) \geq ||\mathbf{T}(t)|| \cos \theta
\end{align}where, $\hat{\mathbf{n}} \in \mathbb{R}^3$ is a unit vector along the axial direction of the spacecraft and pointing down, and $\theta \in [0,\pi]$ is a fixed pre-specified maximum angle between the thrust vector $\vT(t)$ and $\hat{\mathbf{n}}$. Intuitively, this constraint is required for sensors such as cameras to ensure that the ground is always in the field-of-view. For values of $\theta>\pi/2$, this also leads to non-convexity which our proposed method can handle. However, to ensure practical usefulness of maintaining the ground in the field-of-view, we assume $\theta < \pi/2$.
\par Next we introduce state constraints commonly considered in PDG problems to ensure a soft-landing at a pre-specified landing zone. Our strategy is to handle these as soft constraints and penalize violations. In what follows, we will introduce and add terms to our terminal and running cost functions that are used in our stochastic optimal control algorithm. The goal of the algorithm is to minimize the expected running and terminal costs, where the expectation is evaluated using trajectories sampled according to \eqref{eqn:sde_dynamics}. Similar to \cite{exarchos2019optimal}, because the approach discussed in this paper requires trajectory sampling, it is imperative to impose an upper bound on the duration of each trajectory. This is because it is possible to encounter trajectory samples with very large or infinite duration that cannot be simulated. Moreover, it is practically meaningless to continue the simulation if a landing or crash occurs prior to reaching this upper bound. Thus, we formulate a \textit{first-exit} problem with a finite upper bound on the time duration where the simulation is terminated when one of the following two conditions is met: 1) we reach the ground, i.e., $r_3 = 0$ (or more realistically some threshold $r_3 \leq h_\text{tol}$ where $h_\text{tol}$ is some arbitrarily small number defining a height at which shutting off the thrusters would be considered safe), or 2) the time elapsed during simulation is equal or greater to a predetermined maximum simulation time ($t_f$ seconds), whichever occurs first. Mathematically, the \textit{first-exit} time, $\mathcal{T}$, is defined as follows, 
\begin{align}
    \tau &=\underset{s}{\inf}\big\{s\in[0,\,t_f]\big|\, r_3(s)\leq h_\text{tol}\big\}\nonumber     \\
    \mathcal{T} &= \min(\tau,\,t_f).
\end{align}
\par The vehicle is required to perform a safe landing which is characterized by a zero terminal velocity at a predetermined landing zone. However, in a stochastic setting, the probability of a continuous random variable being exactly equal to a specific value is zero. Thus, under stochastic disturbances, it is unrealistic to impose exact terminal conditions. Our strategy is to penalize the mean-squared deviations from the desired positions and velocities at $t=\mathcal{T}$ seconds and thus approach the target positions and velocities on average. As will be later shown, our simulations demonstrate controlled trajectories that terminate in the vicinity of the desired terminal conditions.  We define the following components of our proposed terminal cost function, 
\begin{enumerate}
    \item $\phi_x = \big(r_1(\mathcal{T})\big)^2$ and $\phi_y = \big(r_2(\mathcal{T})\big)^2$,
    where, without loss of generality, we consider the $x$ and $y$ coordinates of the landing zone to be at the origin.
    \item $\phi_z = \big(r_3(\mathcal{T})\big)^2$, where,  we penalize the residual altitude at $t=\mathcal{T}$ seconds to discourage hovering. % as well as going underground i.e. resulting in a crash.
    \item $\phi_{v_x} = \big(\dot{r}_1(\mathcal{T})\big)^2$ and $\phi_{v_y} = \big(\dot{r}_2(\mathcal{T})\big)^2$, where, we penalize the residual $x$ and $y$ velocities at $t=\mathcal{T}$ seconds to discourage tipping over.
    \item $\phi_{v_z} = \begin{cases} c_{v_z +} \big(\dot{r}_3(\mathcal{T})\big)^2, \quad \dot{r}_3(\mathcal{T}) > 0 \text{ m/s}  \\  c_{v_z -} \big(\dot{r}_3(\mathcal{T})\big)^2, \quad \dot{r}_3(\mathcal{T}) \leq 0 \text{ m/s} \end{cases} $ \\namely residual vertical velocity terms with constants $c_{v_z +}$ and  $c_{v_z -}$, where positive terminal velocities are penalized higher by setting $c_{v_z +} > c_{v_z -}$ in order to discourage hovering around the landing zone.
\end{enumerate}
An inequality constraint on the spacecraft's total mass given by, $m(\mathcal{T}) \geq m_d$, is commonly used to ensure that the dry mass ($m_d$ kgs) of the vehicle is lower than the total mass at terminal time $\big(m(\mathcal{T})\big)$. We enforce this constraint as follows:
\begin{equation*}
    \phi_m = \exp{\bigg(-\dfrac{m(\mathcal{T})-m_d}{m(0)-m_d}\bigg)}
\end{equation*}
wherein, the penalty increases exponentially if the terminal mass $(m(\mathcal{T}))$ falls below the dry mass $m_d$. Additionally, this also encourages minimum fuel consumption as higher values of $\big(m(\mathcal{T})-m_d\big)$ lead to lower values of $\phi_m$. 
\par The terminal cost function can now be stated as a weighted sum of the terms described above,\begin{equation}
\label{eqn:terminal_cost}
    \phi(\vx(\mathcal{T})) = Q_x\cdot \phi_x + Q_y\cdot \phi_y + Q_z\cdot \phi_z + Q_{v_x}\cdot \phi_{v_x} + Q_{v_y}\cdot \phi_{v_y} + Q_{v_z}\cdot \phi_{v_z} + Q_{m}\cdot \phi_{m}
\end{equation} where, the coefficients ($Q_i$) allow to tune the relative importance of each term in the terminal cost function. 
\par A glide-slope constraint is also commonly employed to keep the vehicle in an inverted cone with the tip of the cone at the landing zone \cite{dueri2017customized}.  This is given by, 
\begin{equation}
\label{eqn:glideslope_constraint}
    \tan\gamma\cdot\Big|\Big|\big(r_1(t),\,r_2(t)\big)\Big|\Big|\leq r_3(t),
\end{equation} where $\gamma \in [0,\,\pi/2)$ is the minimum admissible glideslope angle. Since, this constraint is imposed at every point in time, we use the following as our running cost funtion,
\begin{align}
    \label{eqn:glide_slope_constraint_cost}
    \Delta_\text{glide} &= \tan{\gamma} \cdot \sqrt{r_1(t)^2 + r_2(t)^2} - r_3(t)\\
    \nonumber
    l(t,\,\vx(t)) &= \begin{cases} q_+ \cdot \Delta_\text{glide}^2 ,\quad \Delta_\text{glide}>0 \\ q_-  \cdot \Delta_\text{glide}^2, \quad  \Delta_\text{glide} \leq 0 \end{cases} \text{where, $q_+ >> q_-$ to penalize trajectories from leaving the glide-slope cone}
\end{align}
Note that we do not set $q_-$ to zero as this encourages hovering around the landing zone at high altitudes by making $\Delta_\text{glide}$ highly negative. Thus, a non-zero value for $q_-$ encourages landing. 
\par Finally, concerning the initial conditions, our formulation allows for $\vx(0) = \big[\vr_0,\,\mathbf{v}_0,\, m_0\big]\T$ to be fixed or sampled from an initial distribution. In our simulations, we train a policy that is able to handle a range of initial positions $\vr_0$ with fixed values of $\mathbf{v_0}$ and $m_0$. This is justified as follows: we assume that separate navigation systems onboard the spacecraft take care of the main flight segment (e.g., from planet to planet) and will navigate the spacecraft to a position that is within reasonable distance from the landing zone for the final descent stage to begin. Specifically, we assume that the final descent stage is initialized when the spacecraft reaches a certain altitude. As far as the corresponding initial $x$,$y$ coordinates are concerned, we assume that these lie on the base of an inverted cone as defined by \eqref{eqn:glideslope_constraint}. The radius depends on the accuracy we expect to see from the main navigation system: the higher its accuracy, the closer the initial $x$,$y$ positioning will be to the landing zone, though in any case the exact values will not be known to us \textit{a priori}.

\subsection{The Minimum Fuel or $\mathcal{L}^1$ Stochastic Optimal Control Problem}
We can now formulate the PDG stochastic optimal control problem as a constrained non-convex minimization problem where the goal is to minimize the amount of fuel needed to achieve a safe landing. As motivated in the introduction and in \cite{imross2004} we consider the $\mathcal{L}^1$-norm of the thrust as the running control cost (as opposed to the conventional quadratic cost or $\mathcal{L}^2$-norm) to correctly measure and hence minimize the total fuel consumption. The optimization problem is formally stated as,

\noindent 
minimize: \quad \quad \quad $J\big(t=0,\, \vx(t),\, \vT(t)\big) =  \Eb_\Qb \Bigg[ \phi\big(x(\mathcal{T})\big) + \displaystyle\int_{0}^{\mathcal{T}} \bigg( l\big(s,\,\vx(s)\big) + q \big(||\vT(t)||_{\mathcal{L}^1}\big)\bigg)\rd s \Bigg] $ \\
\noindent
subject to:
\begin{align}
\label{eqn:general_formulation}
\begin{split}
    &\rd r(t) = \rd\mathbf{v}(t)\rd t, \\ 
    &\rd\mathbf{v}(t) =  \frac{\textbf{T}(t)}{m(t)}\rd t - \textbf{g}\rd t + \frac{\Gamma}{m(t)} \rd W(t), \\ 
    &\rd m(t) = - \alpha \bigg [ ||\vT(t)||_2\rd t + \mathbf{1}_{1 \times 3}\T \Gamma \rd W(t) \bigg ], \\
    &0 < \rho_1 \leq ||\mathbf{T}(t)||_2 \leq \rho_2, \quad \hat{\mathbf{n}}\cdot \mathbf{T}(t) \geq ||\vT(t)||_2 \cos \theta \\
\end{split}
\end{align}
where, $\phi: \mathbb{R}^n \to \mathbb{R}^+$ is defined as per eqn. \eqref{eqn:terminal_cost}, $l: \mathbb{R}^n \to \mathbb{R}^+$ is defined as per eqn. \eqref{eqn:glide_slope_constraint_cost}, and $q$ is a positive scalar weight assigned to the $\mathcal{L}^1$-norm of the thrust vector. 
\par There are three sources of nonconvexity in the presented problem formulaton,
    \begin{enumerate}
        \item the relationship between the mass-rate $\big(\dot{m}(t)\big)$ and the thrust vector $\big(\vT(t)\big)$ in the dynamics,
        \item the lower bound on the norm of the thrust vector $\big(\rho_1 \leq ||\vT(t)||_2\big)$, and,
        \item the thrust-pointing constraint when $\theta > \pi/2$
    \end{enumerate}
Existing work in literature \cite{dueri2017customized, reynolds2020real} either attempt to convexify the original problem and then use customized convex solvers or rely on sequential convex programming to iteratively convexify and solve the original nonlinear problem. In contrast to these methods, our approach can handle the nonlinear dynamics and does not require any convexification.
\subsection{Solution using Forward and Backward Stochastic Differential Equations}
In this section, we describe our methodology to solve the $\mathcal{L}^1$ stochastic optimal control problem described in equation  \eqref{eqn:general_formulation}. We seek to minimize the expected cost with respect to the set of all admissible controls $\mathcal{U}$. We begin by defining the value function ($V$) (i.e., the \textit{minimum cost-to-go}) as follows, \begin{align}
\begin{cases}
    V\big(\vx(t),\, t\big) = \inf_{\vT(\cdot) \in \mathcal{U}[0, \mathcal{T}]}  J\big(t=0,\, \vx(t),\,\vT(t)\big) \\
    V\big(\vx(\mathcal{T}),\,  \mathcal{T}\big) = \phi\big(\vx(\mathcal{T}),\, \mathcal{T}\big)
\end{cases}
\end{align}Using Bellman's principle of optimality and applying Ito's lemma, one can derive the HJB-PDE given by,\begin{align}
\begin{cases}
\label{eqn:hjb_pde_general}
    V_t + \inf_{\vT(\cdot) \in \mathcal{U}[0, \mathcal{T}]}  \bigg  \{ \frac{1}{2} tr \big( V_{\vx\vx} \Sigma \Sigma \T \big) + V_\vx \T f\big(\mathbf{x}(t),\, \vT(t),\, t\big) + l\big(\mathbf{x}
    (t),\, t\big) + q \big| \big|\vT(t) \big| \big|_{\mathcal{L}^1} \bigg \} = 0 \\
    V(\mathbf{x}(\mathcal{T}),\mathcal{T}) = \phi(\mathbf{x}(\mathcal{T}), \mathcal{T})
\end{cases}
\end{align}where the subscripts $t$ and $\vx$ are used to denote partial derivatives with respect to time and state, respectively. The term inside the infimum operator is known as the Hamiltonian (denoted $\mathcal{H}$). The HJB-PDE is a backward, nonlinear parabolic PDE and solving it using grid-based methods is known to suffer from the well-known \textit{curse-of-dimensionality}. 
Among some of the recent scalable methods to solve nonlinear parabolic PDEs, the Deep FBSDEs \cite{pereira2019learning, pereira2020feynman, pereira2020safe} based solution is the most promising and has been used successfully for high-dimensional problems in finance \cite{exarchos2020novas}. Deep FBSDEs leverage the function approximation capabilities of deep neural networks to solve systems of FBSDEs which in turn solve the corresponding nonlinear parabolic PDE. The connection between the soluions of nonlinear parabolic PDEs and FBSDEs is established via the nonlinear Feynman-Kac lemma \cite[Lemma 2]{exarchos2018stochastic}. Thus, applying the nonlinear Feynman-Kac lemma yields the following system of FBSDEs,
\begin{align}
    \label{eqn:fsde}
    \vx(t) &= \vx(0) + \int_0^t f\big(\vx(t),\,\vT^*(t),\,t)\big) \,\rd t + \int_0^t \Sigma\big(\vx(t),\,t)\big) \,\rd W(t)\quad \text{[FSDE]}\\
    \label{eqn:bsde}
    V\big(\vx(t),\,t\big) &= \phi\big(\vx(\mathcal{T})\big)+\int_t^\mathcal{T} \bigg( l\big(\mathbf{x}
    (t),\, t\big) + q \big| \big|\vT^*(t) \big| \big|\bigg) \rd t -  \int_t^\mathcal{T} V_\vx\T\Sigma\big(\vx(t),\,t)\big)\,\rd W(t) \quad \text{[BSDE]}\\
    \label{eqn:hamiltonian_minimization}
    \vT^*(t) &= \underset{\vT \in \mathcal{U}}{\text{argmin}}\; \mathcal{H}\big(\vx(t),\,\vT(t),\,V_\vx,\,V_{\vx\vx}\Sigma\Sigma\T\big)\quad \text{[Hamiltonian minimization]}
\end{align}
Because of the terminal condition $\phi\big(\vx(\mathcal{T})\big)$,  $V\big(\vx(t),\,t\big)$ evolves backward in time while $\vx(t)$ evolves forward in time yielding a two-point boundary value problem. Thus, simulating $\vx(t)$ might be trivial, however $V\big(\vx(t),\,t\big)$ cannot be naively simulated by backward integration like an ODE. This is because within the Ito integration framework, in order for solutions to be adapted, the process should be non-anticipating; which means that in this case naive backward integration of $V\big(\vx(t),\,t\big)$  would result in it depending explicitly
on future values of noise making it an anticipating stochastic process. 
 One solution to solve BSDEs is to backward-propagate the conditional expectation of the process as was done in \cite{exarchos2018stochastic}. However, the least-squares-based algorithm to approximate the conditional expectation suffers from compounding approximation errors at every time step and thus cannot scale. To overcome this, the deep FBSDE method \cite{pereira2019learning} parameterizes the unknown value function $V\big(\vx(0),\,0;\,\theta\big)$ and the gradient of the value function $V_{\vx}\big(\vx(t),\,t;\,\theta\big)$ using an LSTM-based deep neural network. The parameters $\theta$ of the network are trained using Adam \cite{kingma2014adam} or any variant of the stochastic gradient descent algorithm. By introducing an initial condition, the BSDE is forward propagated as if it were a forward SDE and the known terminal condition $\bigg(V\big(\vx(\mathcal{T}),\,\mathcal{T}\big)=\phi\big(\vx(\mathcal{T})\big)\bigg)$ is used as a training loss for the deep neural network. This solution has been demonstrated to be immune to compounding errors and can scale to high-dimensional problems \cite{pereira2019learning, pereira2020feynman, pereira2020safe}. The Hamiltonian minimization at every time step computes the optimal control (i.e., the optimal thrust) that is used in the drifts of the FSDE and the BSDE. For numerical simulations, the system of FSBDEs is discretized in time using an Euler-Maruyama discretization to yield the following set of equations,
 \begin{align}
 \label{eqn:discrete_fbsde}
     \vx[k+1] &= \vx[k] + f\big(\vx[k],\,\vT^*[k],\,k)\big) \,\Delta t + \Sigma\big(\vx[k],\,k)\big) \,\Delta W[k]\\
     V\big(\vx[k+1],\,k+1\big) &= V\big(\vx[k],\,k\big)+ l\big(\mathbf{x}
    [k],\, k\big) \Delta t + q \big| \big|\vT^*[k] \big| \big| \Delta t - V_\vx\T\Sigma\big(\vx[k],\,k)\big)\,\Delta W[k]\\
    \vT^*[k] &= \underset{\vT \in \mathcal{U}}{\text{argmin}}\; \mathcal{H}\big(\vx[k],\,\vT[k],\,V_\vx,\,V_{\vx\vx}\Sigma\Sigma\T\big)
 \end{align}
 where $k$ denotes the discrete-time index and $\Delta t$ denotes the time-interval (in continuous-time) between any two discrete-time indices $k$ and $k+1$. 
 \par For systems with control-affine dynamics and quadratic running control costs (or $\mathcal{L}^2$ norm of control) as in \cite{pereira2019learning}, this minimization step has a closed form expression. For the one dimensional soft-landing problem as in \cite{exarchos2019optimal}, the closed-form expression yields the well-known \textit{bang-bang} optimal control solution due to presence of the $\mathcal{L}^1$ norm in the running control cost. However, for the general soft-landing problem in three dimensions, as presented in this paper, the dynamics are non-affine with respect to the controls. As a result, a closed-form \textit{bang-bang} optimal control cannot be derived and the Hamiltonian minimization step requires a numerical solution. 
Additionally, as described in equation \eqref{eqn:general_formulation}, the general problem  has non-trivial control constraints with non-affine dynamics. In the following section, we build off of recent work \cite{exarchos2020novas} that embeds a non-convex optimizer into the deep FBSDE framework to solve non-convex Hamiltonian minimization problems at each time step. We extend this framework to handle the aforementioned control constraints as well as the first-exit problem formulation. Moreover, as stated in \cite{exarchos2020novas} this non-convex optimizer is differentiable and can
facilitate end-to-end learning making it a good fit to be embedded within the deep FBSDE framework.

\section{Proposed Solution using  NOVAS-FBSDE}
The presence of $||\cdot||_2$ in the equation for $\dot{m}(t)$ makes the dynamics a non-affine function of the control, $\vT(t)$. Additionally, the control constraints given by equations \eqref{eqn:thrust_bounds} and \eqref{eqn:thrust_pointing} are non-convex as described in previous sections. As a result, 
the Hamiltonian minimization at each time step is a non-convex optimization problem. The general Hamiltonian ($\mathcal{H}$) takes the following form,  
\par(Note: henceforth the dependence of $V_\vx,\,V_{\vx\vx}$ and $\Sigma$ on $\vx$ and $t$ will be dropped for ease of readability)
$$\mathcal{H} \big(\vx(t),\,\vT(t),\,V_\vx,\,V_{\vx\vx}\Sigma\Sigma\T\big) \triangleq \frac{1}{2}tr\big(V_{\vx\vx}\Sigma\Sigma\T\big) + V_\vx \T f\big(\vx(t), \vT(t)\big) + l\big(t, \vx(t), \vT(t)\big)$$
However, in this problem, the diffusion matrix $\Sigma$ is not dependent on the control $\vT(t)$ i.e., we do not consider control-multiplicative noise entering the dynamics. As a result, the trace-term can be ignored from the above expression and unlike \cite{exarchos2020novas} we do not require an extra neural network to predict the terms of the hessian of the value function $V_{\vx\vx}$. Thus, the simplified Hamiltonian for our problem that ignores terms not dependent on $\vT(t)$ is given by, \begin{equation}
\label{eqn:simplified_hamiltonian}
    \mathcal{H} \big(\vx(t),\,\vT(t),\,V_\vx\big)=V_\vx \T f\big(\vx(t), \vT(t)\big) + q\big|\big|\vT(t)\big|\big|_{\mathcal{L}^1}
\end{equation}
\par To handle non-convex Hamiltonian minimization within deep FBSDEs, recently, a new framework \cite{exarchos2020novas} was developed that combines deep FBSDEs with the Adaptive Stochastic Search algorithm \cite{zhou2014gradient} to solve such problems while allowing efficient backpropagation of gradients to train the deep FBSDE network. This framework is called NOVAS-FBSDE wherein NOVAS stands for \textit{Non-Convex Optimization Via Adaptive Stochastic Search}. NOVAS has been demonstrated to recover the closed-form optimal control in case of control-affine dynamics and has been tested on high-dimensional systems such as portfolio optimization with 100 stocks \cite{exarchos2020novas} in simulation. In a nutshell, at each time step, the Hamiltonian ($\mathcal{H}$) is minimized using the Adaptive Stochastic Search (GASS) algorithm. Briefly stated, Adaptive Stochastic Search first converts the original deterministic problem into a stochastic problem by introducing a parameterized distribution $\rho(\vT(t);\,\theta)$ on the control $\vT(t)$ and shifts the minimization of $\mathcal{H}$ with respect to $\vT(t)$ to minimization of $\mathbb{E}[\mathcal{H}]$ with respect to $\theta$. This allows for $\mathcal{H}$ to be an arbitrary function of $\vT(t)$ (potentially non-differentiable) and $\mathbb{E}[\mathcal{H}]$ is approximated by sampling from $\rho(\vT(t);\,\theta)$. By minimizing $\mathbb{E}[\mathcal{H}]$, the upper bound on $\mathcal{H}$ is minimized. We invite the reader to refer to appendix \ref{sec:novas_derivation} for a detailed exposition of the equations in NOVAS and its algorithmic details.
\par Notice that the general problem \eqref{eqn:general_formulation} has hard control constraints (i.e. equations \eqref{eqn:thrust_bounds} and \eqref{eqn:thrust_pointing}). To enforce these constraints, we employ a novel sampling scheme based on the lemma given below. We make the following assumptions,
\begin{definition}

\label{assumption:1}
 The horizontal thrust components $\big(\vT_1(t),\,\vT_2(t)\big)$ are bounded based on the lower bound of the norm of the thrust $\rho_1$, so that $|\vT_1(t)|\leq \dfrac{\rho_1}{2}$ and $|\vT_2(t)|\leq \dfrac{\rho_1}{2}$.
\end{definition}

\begin{definition}
\label{assumption:2}
    The bounds on the norm of the thrust vector $\vT(t)$ are such that $0<\rho_1<<\rho_2$.
\end{definition}

\begin{definition}
\label{assumption:3}
     The maximum angle $\theta$ between the thrust vector $\vT(t)$ and $\hat{\vn}$ belongs to the interval  $\bigg[\dfrac{\pi}{6}, \dfrac{\pi}{2}\bigg)$.
\end{definition}

\begin{definition}
\label{assumption:4}
    The bounds $\rho_1, \, \rho_2$ and the angle $\theta$ satisfy, $\sqrt{\dfrac{\rho_1^2}{2\cdot \sin^2{\theta}}} \leq ||\vT(t)|| \leq \rho_2 $.
\end{definition}

The assumption \ref{assumption:3} is justified because values of $\theta\geq \pi/2$ will result in the camera sensors loosing the ground from their field of view, while very low values of $\theta$ will restrict horizontal motion. 
\par Since, $\hat{\mathbf{n}}=[0,\,0,\,1]^\mathrm{T}$, the \textit{thrust-pointing} control constraint that must be satisfied is $\hat{\mathbf{n}}\cdot \vT = \vT_3\geq ||\vT||\cos{\theta}$. 
\begin{lemma}
\label{lemma:thrust_pointing}
 Given that assumptions \ref{assumption:1}$-$ \ref{assumption:4} hold, the \textit{thrust-pointing} constraint $\vT_3\geq ||\vT||\cos{\theta}$ is satisfied. 
\end{lemma}
\renewcommand\qedsymbol{$\blacksquare$}
\begin{proof}
Given that,  $\sqrt{\dfrac{\rho_1^2}{2\cdot \sin^2{\theta}}} \leq ||\vT(t)|| \leq \rho_2$, we have $\dfrac{\rho_1^2}{2\cdot \sin^2{\theta}} \leq ||\vT(t)||^2 \leq \rho_2^2$.
\\\par  $\therefore \rho_1^2 \leq ||\vT||^2\, \sin^2{\theta} = ||\vT||^2 (1 - \cos^2{\theta}) = ||\vT||^2 - ||\vT||^2\,\cos^2{\theta}$
\par Based on assumption \ref{assumption:1}, we have $\vT_1^2 + \vT_2^2 \leq \dfrac{\rho_1^2}{4} < \rho_1^2$. Therefore the above inequality becomes, 
\par $\vT_1^2 + \vT_2^2 \leq ||\vT||^2 - ||\vT||^2\,\cos^2{\theta}$
\par $\therefore ||\vT||^2\,\cos^2{\theta} \leq ||\vT||^2 -\vT_1^2 - \vT_2^2 = \vT_3^2$
\par $\implies ||\vT||\,\cos{\theta} \leq \vT_3$
\end{proof}
Thus, for lemma \ref{lemma:thrust_pointing} to hold, we need to satisfy assumptions \ref{assumption:1}$-$ \ref{assumption:4}. Assumptions \ref{assumption:2} and \ref{assumption:3} are satisfied by design decisions. For assumptions \ref{assumption:1} and \ref{assumption:4} we sample the horizontal thrust components  $\big(\vT_1(t),\,\vT_2(t)\big)$ and the norm of the thrust $||\vT(t)||=\sqrt{\vT_1^2(t) + \vT_2^2(t) + \vT_3^2(t)}$ and project these samples onto closed intervals such that both assumptions along with the original thrust bounds of eqn. \eqref{eqn:thrust_bounds} are satisfied. Defining $\rho_3 = \sqrt{\dfrac{\rho_1^2}{2\cdot \sin^2{\theta}}}$ and projecting the samples of $||\vT(t)||$ onto the interval $\big[\max(\rho_1,\,\rho_3), \, \rho_2\big]$, both control constraints (equations \eqref{eqn:thrust_bounds} and \eqref{eqn:thrust_pointing}) can be satisfied. A pseudo-code of this sampling scheme is presented in the appendix Algorithm \ref{alg:sample_controls}.

\section{Algorithmic Details}
\label{section:alg}

In this section we present algorithmic details concerning (a) sampling for control constraints, (b) training of the NOVAS-FBSDE network with \textit{first-exit} times and (c) the capability to handle random initial starting positions, which differentiate the proposed framework from algorithms presented \cite{pereira2019learning} and \cite{exarchos2020novas}. A diagram incorporating architectural changes of the deep neural network to enable these new capabilities is also presented.

\subsection{NOVAS with control constraints}
The pseudo-code in Alg. \ref{alg:sample_controls} details the sampling procedure to enforce control constraints at each time step within the NOVAS module of the NOVAS-FBSDE architecture. Similar to \cite{exarchos2020novas}, to sample controls we assign a univariate Gaussian distribution to each control dimension and optimize the parameters of each Gaussian. Thus, the inputs to the NOVAS sampling module are the mean and standard deviation for the lateral thrust components and the mean and standard deviation for the norm of the thrust vector. Before the samples are evaluated to compute the control update, each sample is projected onto a closed interval to satisfy the aforementioned hard control constraints. From numerous experiments, we observed that warm-starting the NOVAS module by using the optimal control from the previous time step as the initial mean, resulted in temporally coherent and less noisy control trajectories. Additionally, it allows using fewer inner-loop iterations within NOVAS.

\subsection{Deep FBSDEs with first-exit times}
So far deep FBSDEs have been successfully implemented for fixed finite time-horizon problems (i.e., $\mathcal{T}=t_f$ is constant). In order to incorporate first-exit times as required in our problem formulation, we use a mask such that,

$$\text{mask}=\begin{cases} 1, \quad r_3(t) > h_\text{tol} \\ 0, \quad r_3 \leq h_\text{tol} \end{cases}$$
where, $h_\text{tol}>0$ m, is a user-defined fixed tolerance for the altitude to determine if a landing (or a crash) or the maximum simulation time (i.e., first-exit) has occurred. In the deep FBSDE framework, multiple (i.e., a mini-batch) trajectories are simulated in parallel in order to train the network with the Adam optimizer \cite{kingma2014adam}. Thus, due to stochastic dynamics, each trajectory could potentially have a different first-exit time. To keep track of these different first-exit times, we maintain a vector of masks of the same size as the mini-batch which is then incorporated into the equations of the forward and backward SDEs. The pseudo-code (Alg. \ref{alg:first_exit}) provides further details regarding the forward pass of the NOVAS-FBSDE architecture. The forward pass ends once all trajectories have been propagated to a maximum time step of $t=t_f$ seconds. 
%To reiterate, first-exit either occurs when the spacecraft has crashed or landed before maximum time $T$ occurs or if maximum time $T$ has been reached and the spacecraft has not landed. 
If first-exit does occur before $t_f$ seconds, the dynamics are "frozen" and propagated until $t=t_f$ using an identity map. This allows to use the same trajectory length for all batch elements, use the terminal state (rather than first-exit state) from all batch elements to compute a loss and to back-propagate gradients during the backward pass up to the time step of first-exit in order to minimize the loss incurred at first-exit time. The output of the forward pass is the $\mathcal{L}oss$ function as shown in Alg. \ref{alg:first_exit} which is then fed to the Adam optimizer to train the NOVAS-FBSDE LSTM network.
\par The Alg. \ref{alg:first_exit} also contains discretized equations of the FSDE and the BSDE. The discretization interval ($\Delta t$ seconds) is fixed and is user-defined. The total number of time steps (or discrete time intervals) is computed as $N=t_f/\Delta t$ such that when $t=t_f$ seconds, the discrete-time index $k=N$ where $ k \in\{0,\,1,\,\ldots,\,N\}$.

% Shall we make V_x an explicit input to the NOVAS function?
% Should we add an algorithm describing how NOVAS performs the minimization? (i.e.: show sampling, computation of the hamiltonian, exponential averaging, etc)
\begin{algorithm}[h!]
\caption{NOVAS-FBSDE with first-exit times}
\label{alg:first_exit}
\begin{algorithmic}[1]
\Function{forward\_pass}{number of time steps ($N$), altitude threshold ($h_\text{tol}$), batch size ($B$), time discretization ($\Delta t$), LSTM neural-network to predict $V_\vx$  ($f_\text{LSTM}$), diffusion matrix ($\Sigma$), system drift $(f)$, Hamiltonian function $(\mathcal{H})$, running cost $(l)$,  inputs and hyperparameters for NOVAS module ($NOVAS\_inputs$), initial value function network ($f_{V_0}$), networks to predict initial LSTM-states ($f_{c_0}^i,\,f_{h_0}^i$), number of LSTM hidden layers ($H$), radius of base of glide-slope cone ($rad$), initial-state vector with uninitialized starting positions $\vx[0]$} 
\State \textbf{Initialize}: \textbf{mask}$\leftarrow 0_{B\times
1}$, $\vx[b=1:B,\,t=0,\,r_1,\, r_2] \leftarrow$ \textbf{sample\_initial\_states} ($B$, $rad$)
\Statex
\Statex \textit{(Predict initial value and the LSTM cell and hidden states)}
\For{$b=1:B$ \textbf{\textit{(in parallel)}}}
 \State $V[b,0]=f_{V_0}(\vx[b,0])$
 \For{$i=1:H$ \textbf{\textit{(in parallel)}}}
    \State $h_i[b,0]=f_{h_0}^i(\vx[b,0])$
    \State $c_i[b,0]=f_{c_0}^i(\vx[b,0])$
 \EndFor
\EndFor
\Statex
\Statex \textit{(Forward propagate the dynamics and value function trajectories)}
\For{$k=0:N-1$}
    \For{$b=1:B$ \textbf{\textit{(in parallel)}}}
    \If{$r_3 >h_\text{tol}$}
        \State \textbf{mask}$[b]\leftarrow 1$
    \Else
        \State \textbf{mask}$[b]\leftarrow 0$
    \EndIf
    \State sample noise, $\Delta w[b,k] \sim 
    \mathcal{N}(\mathbf{0},\,\Delta t \mathbf{I})$
    \Comment{zero mean vector has same dimensionality as $\vx$}
    \State $V_\vx[b,k]\leftarrow f_\text{LSTM}(\vx[b,k])$
    \State $\vT^*[b,k] \leftarrow\,$\textbf{NOVAS\_Layer}$\,(\vx[b,k],\,V_\vx[b,k],\, \mathcal{H}, \, f, \,NOVAS\_inputs)$
    \Statex
    \Comment{warm-start \textbf{NOVAS} with $\vT^*[b,k-1]$ for $k>0$}
    \State \textbf{FSDE:} $\vx[b,k+1] = \vx[b,k] + \textbf{mask}[b]\odot\Big(f\big(\vx[b,k],\, \vT^*[b,k]\big)\,\Delta t + \Sigma\big(\vx[b,k],\, k\big) \,\Delta w[b,k]\Big)$ 
    \State \textbf{BSDE:} $V[b,k+1]=V[b,k] + \textbf{mask}[b] \odot \Big(-l\big(\vx[b,k],\,\vT^*[b,k]\big)\,\Delta t + V_\vx[b,k]\T \Sigma\big(\vx[b,k],\, k\big) \,\Delta w[b,k] \Big)$
    \EndFor
\EndFor
\Statex
\Statex \textit{(Compute loss function using the predicted value and its gradient)}
\State $V^*[N]=\phi\big(\vx[N]\big),\, V_\vx^*[N]=\dfrac{\partial \phi\big(\vx[N]\big)}{\partial \vx},\,V_\vx[N]=f_\text{LSTM}(\vx[N])$\Comment{evaluated in parallel as a batched operation}
\State $\mathcal{L}oss=\dfrac{1}{B}\displaystyle\sum_{b=1}^B \bigg\{\,\big|\big|V[b,N]-V^*[b,N]\big|\big|^2_2 + \big|\big|V_\vx[b,N]-V_\vx^*[b,N]\big|\big|^2_2 + \big|\big|V^*[b,N]\big|\big|^2_2 + \big|\big|V_\vx^*[b,N]\big|\big|^2_2  \,\bigg\}$
\State \Return $\mathcal{L}oss$
\EndFunction
\Statex \begin{itemize}
    \item For $NOVAS\_inputs$, see appendix sections \ref{sec:novas_derivation} and \ref{sec:novas_algo} for the derivation and algorithmic details of the NOVAS module. These are summarized from the work in \cite{exarchos2020novas}.
    \item In steps 20 and 21, for a given batch index $b$, the same \textbf{mask}[$b$] is used for each element of the state vector $\vx$.
    \item The parallel for-loops over $b=1:B$ and $i=1:H$ can be easily implemented with vectorized operations and batched operations, using any deep learning framework such as PyTorch \cite{paszke2019pytorch} or TensorFlow \cite{abadi2016tensorflow}.
\end{itemize}
\end{algorithmic}
\end{algorithm}

\subsection{Training a policy network invariant of initial position}
So far in the deep FBSDEs literature \cite{exarchos2020novas, pereira2019learning, pereira2020safe, wang2019deep, pereira2020feynman} a fixed initial state $\vx[0]$ has been used for every batch index $b$ leading to the network only being able to solve the problem starting from $\vx[0]$. However, this is a very limiting assumption in practice, more so for the planetary soft-landing problem as the probability of the spacecraft being in a specific initial state is zero. To tackle this, we relax this assumption and initialize $\vx$ such that the first two elements i.e., the $x$ and $y$ coordinates of the spacecraft $(r_1[0],\,r_2[0])$ take random values on the base of an inverted cone. This cone corresponds to the glide-slope constraint \eqref{eqn:glideslope_constraint}, with the tip of the cone at the landing zone and the base at some initial altitude $(r_3[0])$. The intuition for uniformly sampling initial positions on the base of this cone is that, in practice, once the spacecraft drops to an altitude of $(r_3[0])$ and is within some radius $rad$ of the landing zone (where $rad$ is the radius of the base of the cone), our trained neural network controller is deployed which then keeps the spacecraft within the glide-slope cone while decelerating towards the landing zone. We provide details regarding sampling initial positions in the pseudo-code Alg. \ref{alg:initial_x0}. A consequence of not starting all batch elements ($b=1:B$) from the same initial starting state $\vx[0]$ is the need to add more neural networks at the initial time step. This is required to approximate the initial value function $(V[0])$ and the initial cell and hidden states of the LSTM neural network (e.g., $(h_0[0], c_0[0])$ and $(h_1[0], c_1[0])$ for a 2-layer LSTM) for each batch element $b$ by feeding these networks with the respective sampled initial positions. These additional networks are shown in Fig \ref{fig:network_architecture}, which is in contrast to all prior deep FBSDE work that only approximates $(V[0])$ with a scalar trainable parameter.

\begin{algorithm}[h!]
\caption{Sampling initial positions on base of glide-slope cone}
\label{alg:initial_x0}
\begin{algorithmic}[1]

\Function{sample\_initial\_states}{$B$, $rad$}       
    \State radii $ = rad\cdot\sqrt{\epsilon_1},\, \epsilon_1 \sim \mathcal{U}(\mathbf{0}_{B\times 1},\,\mathbf{1}_{B\times 1})$ \Comment{sample $B$ uniformly distributed variables}
    \State $\theta = 2\pi \cdot \epsilon_2,\, \epsilon_2 \sim \mathcal{U}(\mathbf{0}_{B\times 1},\,\mathbf{1}_{B\times 1}) $
    \State $r_1 = \text{radii}\cdot \cos(\theta),\, r_2 = \text{radii}\cdot \sin(\theta) $
    \State \Return ($r_1,\,r_2$)
\EndFunction
    \Statex \begin{itemize}
        \item The above square-root, cosine and sine operations are element-wise operations
    \end{itemize}
\end{algorithmic}
\end{algorithm}

\subsection{Network Architecture}
\begin{figure}[ht!]
    \centering
    \includegraphics[scale=0.3]{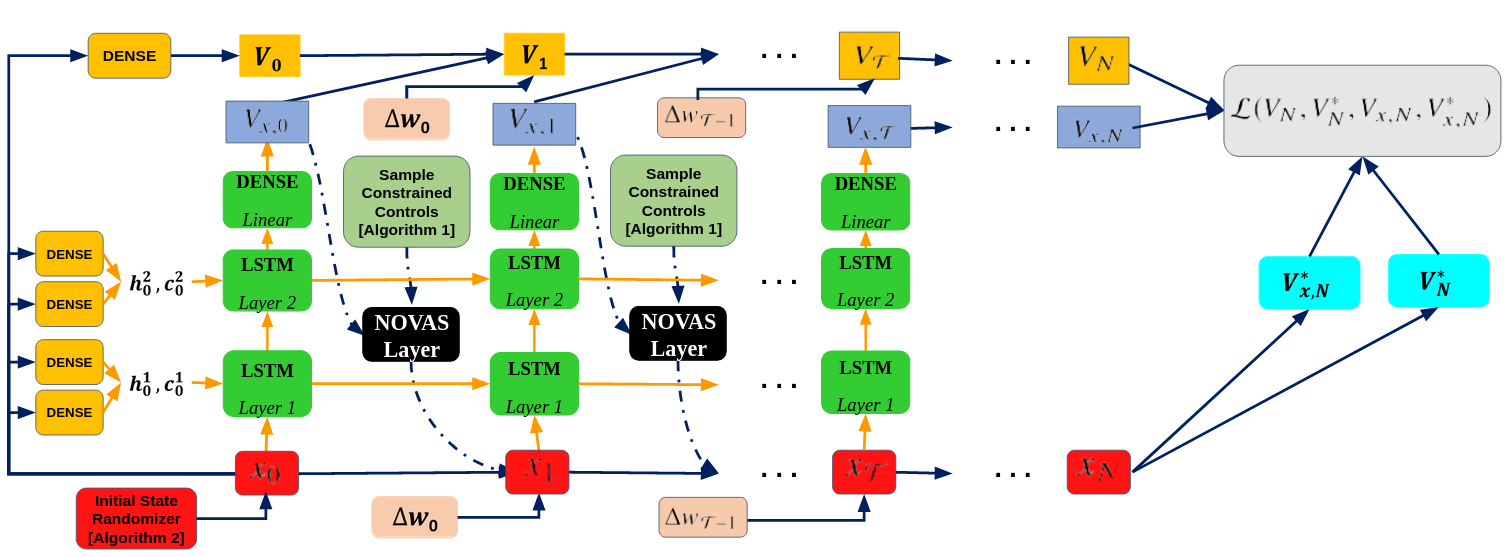}
    \caption{Network Architecture with additional "Dense" i.e. fully-connected layers to enable training from random initial positions. Additionally, the trajectories terminating early at some $\mathcal{T}<t_f$ are frozen using identity maps so that $x_N = x_\mathcal{T}$ allowing gradients to freely flow from time step $N$ to early-exit time step $\mathcal{T}$.}
    \label{fig:network_architecture}
\end{figure}

Similar to past work on deep FBSDEs, we use an LSTM recurrent neural-network architecture to predict the values for the gradient of the value function $V_{\mathbf{x}}(t, \mathbf{x})$ at each time step. These are then used for the computation and minimization, of the Hamiltonian $\mathcal{H}$ inside the NOVAS Layer in fig. \ref{fig:network_architecture}. 

As shown in Fig. \ref{fig:network_architecture}, the random initial position generation algorithm, detailed in Alg \ref{alg:initial_x0}, is used to sample initial positions on the base of the glide-slope cone. This procedure not only makes this approach practically meaningful as discussed in previous sections but also leads to better exploration of the state-space around the landing zone. This was found to significantly improve the performance of the controller when subject to stochasticity in the initial positions and the network can be deployed as a \textit{feedback controller}. We would like to reiterate here that in comparison to existing alternatives although our method requires heavy computation during its training stage, this is not done onboard the spacecraft during the mission. Only the trained policy can be deployed on the spacecraft, and this uses minimal computational resources to predict an optimal control at every time step. The output of Alg. \ref{alg:initial_x0} serves as an input to five two-layer neural networks (with ReLU nonlinearities) whose task is to estimate the initial value of the value function $(V(\vx(0),\,0))$, and the initial values for the hidden and state cells of the two LSTM layers we consider in our architecture. The LSTM layers predict the gradient of the value function, $V_{\mathbf{x}}$, at each timestep which is then used to compute, and minimize, the Hamiltonian at each timestep within the NOVAS layer for a batch of constrained control samples generated by Alg. \ref{alg:sample_controls}. Similar to \cite{pereira2019learning}, the choice of LSTM layers in this architecture is to provide robustness against the vanishing gradient problem, to reduce memory requirements by avoiding individual networks for each time step, to generate a temporally coherent control policy and to avoid the need to feed the time as an explicit input to the network by leveraging the capability of LSTMs to store, read and write from the cell-state (memory). The output of the NOVAS layer is the control (i.e., the thrust) that minimizes the Hamiltonian. This is fed to the dynamical model to propagate forward trajectories until the \textit{first-exit} termination criteria is met. If a particular trajectory is found to terminate early, its state, value function, and gradient of the value function are propagated forward using an identity map for the remaining time steps. This \textit{freezes} the state to the value it takes on at the \textit{first-exit} time. Once the end of the time horizon $t_f$ is reached, we the compute true values for the value function and its gradient using the terminal states. These are fed to a loss function that is used to train the LSTM layers and the neural networks at the initial time step.

\section{Simulation Results} 
% \subsection{3D soft-landing with fixed initial position}
% We first tested our algorithm with a fixed initial position. This was done to primarily  compare the with the results in \cite{dueri2017customized}. Our algorithm is capable of recovering the same results without the need for convexification of the original problem and without the need to resolve the optimization problem at each time step. Additionally, our method can handle stochastic disturbances. Since the work in \cite{dueri2017customized} was performed at NASA's Jet Propulsion Laboratory, the code is not publicly available for comparison.  Therefore, to show comparison we generate the same plots as \cite[Figures 3 and 5]{dueri2017customized} with data generated from our test trials. 
% \subsection{3D soft-landing with random initial positions}
\begin{figure}[ht!]
    \centering
    \includegraphics[scale=0.5]{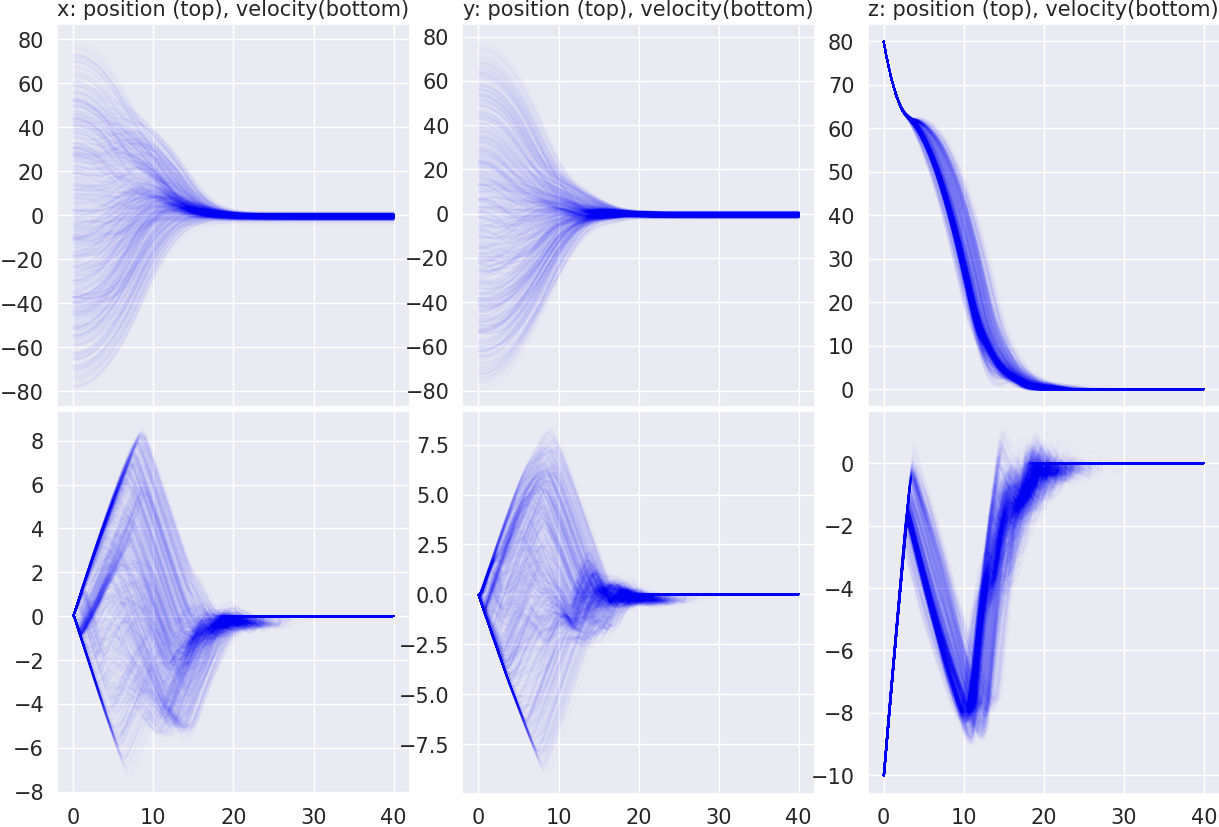}
    \caption{Trajectories of 1024 instances at test-time with 200 NOVAS samples and NOVAS iterations increased to 20 achieving $\mathbf{100.0\%}$ safe landings (note that velocities are zeroed out when a landing or crash is detected).}
    \label{fig:3d}
\end{figure}

% for cone plots 
\begin{figure}[!htb]
\minipage{0.4\textwidth}
  \includegraphics[width=\textwidth]{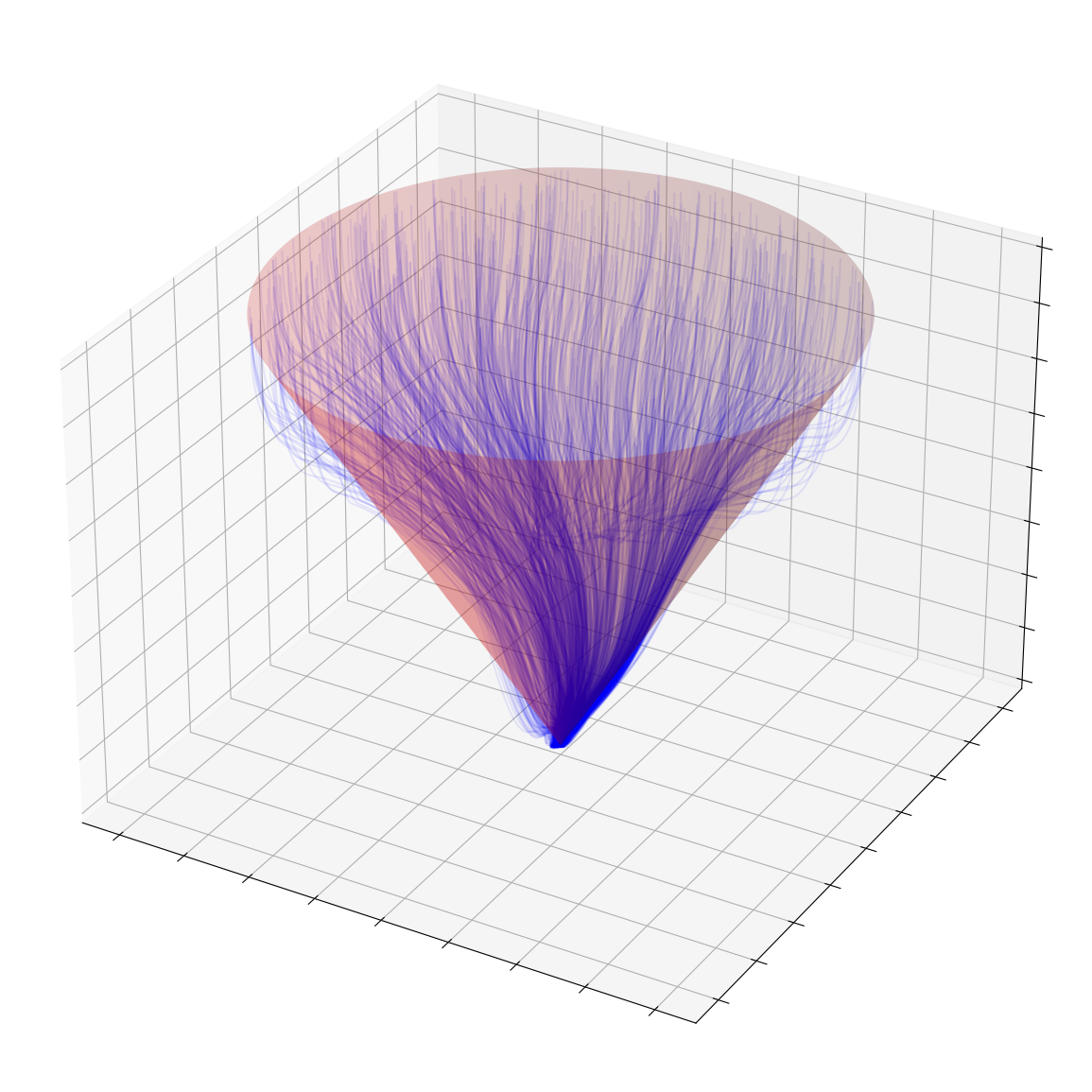}
  \caption{Glide-slope soft constraint: 3D trajectories starting from base of the cone}\label{fig:awesome_image1}
\endminipage
\minipage{0.6\textwidth}
  \includegraphics[width=\textwidth]{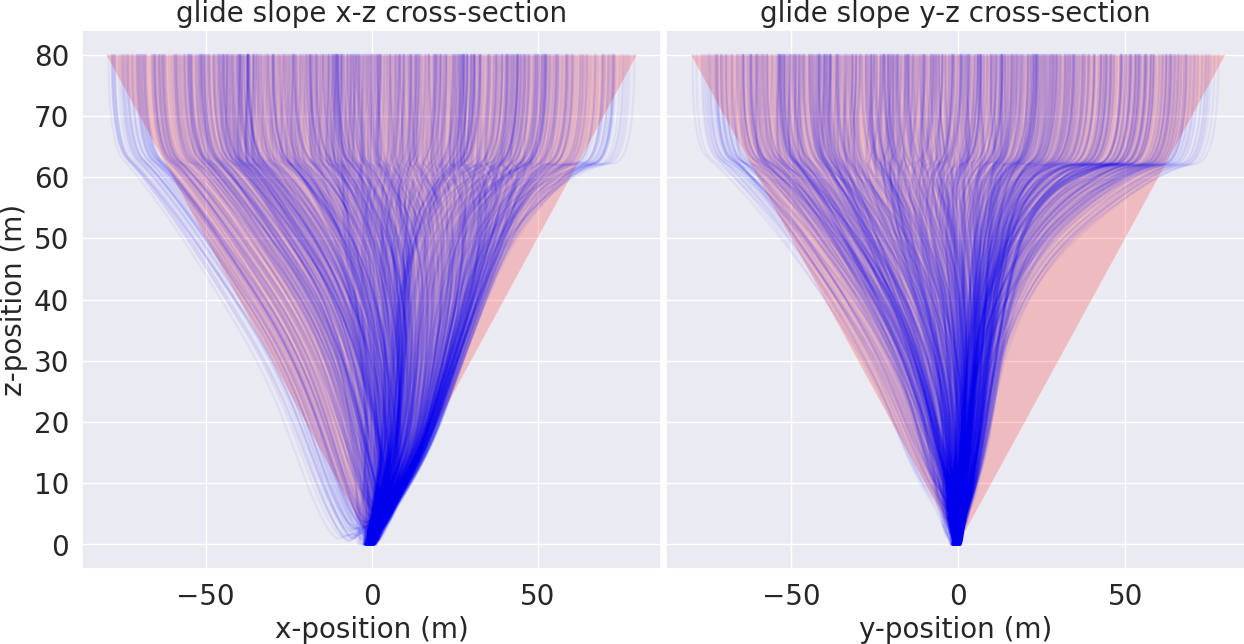}
  \caption{Glide-slope soft-constraint cross-sectional view }\label{fig:awesome_image2}
\endminipage
\end{figure}
\par
\begin{figure}[ht!]
    \centering
    \includegraphics[scale=0.4]{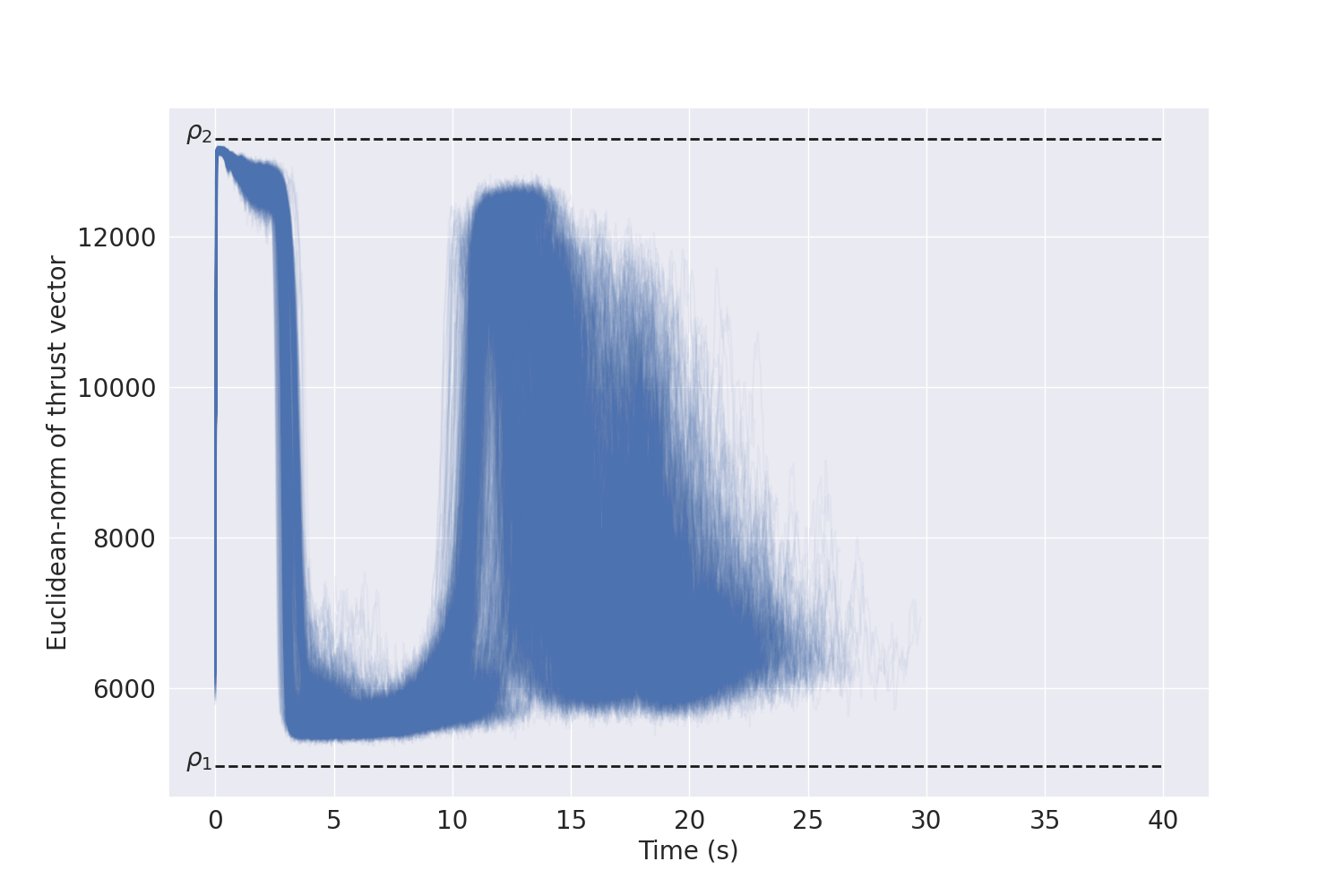}
    \caption{Satisfaction of hard control constraints (note that NOVAS is frozen after landing/crash is detected)}
    \label{fig:control_constraints}
\end{figure}

 We train a NOVAS-FBSDE network for a maximum simulation time of $t_f=20$ seconds and time discretization of $\Delta t = 0.05$ seconds. The network is trained for 7,000 iterations with a learning-rate schedule of $[0.0005,\,0.0001]$, where the learning-rate changes at iteration 3000 from $0.0005$ to $0.0001$. This network is trained with an $\mathcal{L}^1$-norm control cost coefficient of $q=0.00055$. Based on the mass-rate equation for gimbaled rockets \cite{imross2004}, we use the following $\mathcal{L}^1$-norm, $$||\vT(t)||_{\mathcal{L}^1}=\int_0^\mathcal{T}\sqrt{T_1^2(t)+T_2^2(t)+T_3^2(t)}\;\rd t$$ Additionally, we use the following cost coefficients for the terminal cost function, $$Q_x=2.5,\,Q_y=2.5,\,Q_z=2.5,\,Q_{v_x}=5.0,\,Q_{v_y}=5.0,\,Q_{v_z}=10.0,\,Q_{m}=10.0$$  For the altitude tolerance to determine \textit{first-exit} time, we use, $h_\text{tol}=1e^{-3}$. 
Similar to \cite{exarchos2019optimal} we assume a touchdown speed of higher than 5 ft/s\footnote{NASA specifications: \url{ https://www.nasa.gov/mission_pages/station/structure/elements/soyuz/landing.html}} (1.52 m/s) in any direction is considered a crash. We tested with a batch size of 1024 samples and categorized each batch into 3 slots: \textit{not landed, safely landed} and \textit{crashed}. To do this, we use the threshold $h_\text{told}$ to determine if landing has occurred or not and then use the 1.52 m/s threshold to determine if the landing was safe or resulted in a crash. To allow the spacecraft to get close enough to the ground (i.e., below an altitude of $h_\text{tol}$), we increase the maximum simulation time to $t_f=40$ seconds. Note that this is double the maximum simulation time considered during training (i.e., $t_{f,\,\text{training}}=20$ seconds). We hypothesize that because our policy behaves like a \textit{feedback policy}, we can deploy the controller for much longer duration than what it was trained for. We observe the following statistics: $$\text{Not landed}: 1.37\%,\,\text{Safely landed}:98.14\%,\,\text{Crashed}: 0.49\%$$ In order to further improve the test-time results, we increased the number of NOVAS' inner-loop iterations from $10$ iterations used during training to $20$ iterations at test-time. This resulted in $100\%$ safe landings,
$$\text{Not landed}:\mathbf{\color{ForestGreen}{0.0\%}},\,\text{Safely landed}:\mathbf{\color{ForestGreen}{100.0\%}},\,\text{Crashed}: \mathbf{\color{ForestGreen}{0.0\%}}$$
We summarize our observations in Table \ref{table:results}.\begin{table}[h!]
\begin{center}
    \begin{tabular}{||c|c|c|c|c||}
    \hline
        \textbf{NOVAS inner-loop iterations} & \textbf{NOVAS samples} & \textbf{Not landed} & \textbf{Safely landed} & \textbf{Crashed} \\
        \hline\hline
         10 & 200 & 1.37\% & 98.14\% & 0.49\% \\
         15 & 200 & 0.0\% & 99.8\% & 0.2\% \\
         20 & 200 & $\mathbf{\color{ForestGreen}{0.0\%}}$ & $\mathbf{\color{ForestGreen}{100.0\%}}$ & $\mathbf{\color{ForestGreen}{0.0\%}}$ \\
         \hline
    \end{tabular}
    \caption{Landing statistics for 1024 instances with maximum simulation time of $t_{f,\,\text{test}}=40$ seconds}
    \label{table:results}
\end{center}
\end{table} Finally, we demonstrate empirical evidence of satisfaction of hard control constraints by our sampling scheme. For our simulations, we chose $\theta=\pi/4$ which is a reasonable assumption to keep the ground always in the field of view of the camera and other sensors on the base of the spacecraft. Thus, $\rho_3 = \sqrt{\dfrac{\rho_1^2}{2\cdot \sin{^2(\pi/4)}}}=\rho_1$. As seen in fig. \ref{fig:control_constraints} the control-norm \bigg(i.e., $||\vT(t)||_{\mathcal{L}^1}$\bigg) always stays within the bounds of the closed interval $[\rho_1=\rho_3, \,\rho_2]$ and a \textit{max-min-max}-like thrust profile is evident. The controls do not get saturated at the limits because we project the NOVAS samples to the respective feasible sets and then perform gradient updates. Since the updates are convex combinations (due to weights obtained from a \textit{softmax} operation) of samples, the output always lies within the stipulated bounds. For additional details regarding our simulation hyperparameters and computational resources, we invite the reader to refer to sec. \ref{sec:sim_hyperparam} in the appendix. 

\section{Conclusion and future directions}
In this paper, we presented a novel approach to solve the constrained three-dimensional stochastic soft-landing problem using LSTM-based deep recurrent neural networks and the differentiable non-convex optimization layer, NOVAS, within the deep FBSDE framework for end-to-end differentiable $\mathcal{L}^1$ stochastic optimal control. Our approach does not rely on convexification of the constraints or linearization of the dynamics. Through our simulations, we demonstrated empirical satisfaction of hard thrusting (i.e., control) constraints, empirical satisfaction of soft state constraints and empirical robustness to the spacecraft's initial position as well as external disturbances. Our controller is capable of performing safe landings in 93.9\% of the test cases and with additional computation is able to \textbf{safely land all test instances}. Our trained network also exhibits properties of a feedback policy, thereby allowing it to be deployed for a longer duration than the maximum simulation duration during training. Thus, once trained offline, our controller does not require \textit{on-the-go} re-planning as compared to other deterministic methods in literature and can output an optimal control by forward-pass through a neural network and the NOVAS layer. By making the controller robust to the initial position on the base of an inverted cone, not only is the glide-slope of the descent trajectory regulated, but our controller also has a higher tolerance for errors made by the pre-descent stage controllers and can takeover from the previous stage starting in a wide radius around the landing zone. Stemming from these successful results, we propose the following future research paths -  higher dimensional models containing attitude dynamics and constraints on the spacecraft's attitude, risk-sensitive stochastic optimal control for soft-landing, soft-landing-rendezvous problems with a mobile landing platform on land or on water with communication constraints and leveraging NOVAS' ability to handle arbitrary nonlinear dynamics to employ data-driven models such as neural networks to capture phenomena that cannot be easily modeled by explicit equations of motion. 
% \section*{Funding Sources}
\section{Appendix}
\subsection{NOVAS derivation}
\label{sec:novas_derivation}
In this paper, we define the optimization problem so that the optimal control can be computed even in the absence of an analytical solution through the use of the novel approach introduced by Exarchos et. al. in \cite{exarchos2020novas}, by the name of NOVAS. NOVAS stands for \textit{Non-convex Optimization Via Adaptive Stochastic Search}. NOVAS is designed to tackle very general non-convex optimization problems, and is inspired by a well-researched method used across the field of stochastic optimization known as Gradient-based Adaptive Stochastic Search (GASS) \cite{zhou2014gradient}. We summarize here the main ideas, derivation and algorithm from the work \cite{exarchos2020novas} for a quick reference for the reader. For more details and other applications of NOVAS, we invite the interested reader to refer to \cite{exarchos2020novas}. In general, adaptive stochastic search addresses a maximization problem of the following form,\begin{equation}
\label{eqn:gass_problem}
    x^* \in \argmax \limits_{x \in \chi} F(x), \quad \chi \subseteq \mathbb{R}^n
\end{equation} where, $\chi$ is non-empty and compact, and $F: \chi \to \mathbb{R}$ is a real-valued function that may be non-convex, discontinuous and non-differentiable. Given that $F(x)$ is allowed to be very general, this function may be defined by an analytical expression or a neural network. GASS allows us to solve the above maximization problem through a stochastic approximation. For this we first convert the deterministic problem above into a stochastic one in which $x$ is a random variable. Moreover, we assume that $x$ has a probability distribution $f(x;\rho)$ from the exponential family and is parameterized by $\rho$. Using this approximation, we can solve the problem in \eqref{eqn:gass_problem} by solving,\begin{equation}
    \rho^* = \argmax \limits_{\rho} \int F(x) f(x;\rho) \,\rd x = \mathbb{E}_{\rho}[F(x)]
\end{equation}It is common practice to introduce a natural log and a shape function, $S(\cdot)$ with properties of being a continuous, non-negative and non-decreasing function. Due to these properties, the optima of the new problem remain unchanged. The problem then becomes,\begin{equation}
\label{eqn:sample_objective}
    \rho^* = \argmax \limits_{\rho} \ln \int S(F(x)) f(x;\rho) \,\rd x = \ln \mathbb{E}_{\rho}[S(F(x))]
\end{equation}
Notice that the optimization is not with respect to $x$ anymore and is instead with respect to the parameters of the distribution on $x$. Thus, we can attempt to solve the above problem with gradient-based approaches as the non-differentiability with respect to $x$ has now been circumvented. The only difference is that we now optimize for the expected objective and thus a lower bound on the true (local) maximum. Taking the gradient of the objective we have, 
\begin{align*}
    \nabla_\rho \ln \int S(F(x)) \,f(x;\rho) \,\rd x &= \dfrac{\int S(F(x))\, \nabla_\rho f(x;\rho) \,\rd x}{\int S(F(x)) \,f(x;\rho) \,\rd x} \\
    &=\dfrac{\int S(F(x))\, \nabla_\rho f(x;\rho) \,\dfrac{f(x;\rho)}{f(x;\rho)}\rd x}{\int S(F(x)) \,f(x;\rho) \,\rd x} \\
    &=\dfrac{\int S(F(x))\, \nabla_\rho \ln f(x;\rho) \,f(x;\rho)\rd x}{\int S(F(x)) \,f(x;\rho) \,\rd x} \quad \text{(also known as the log-trick)}\\
    &=\dfrac{\mathbb{E}\big[S(F(x))\, \nabla_\rho \ln f(x;\rho)\big]}{\mathbb{E}\big[S(F(x))\big]}
\end{align*}
The \textit{log-trick} allows us to approximate the gradient by sampling. This makes this method amenable to GPUs or vectorized operations. Since $f(x;\rho)$ belongs to the exponential family we can compute an analytical form for the gradient inside the expectation. Distributions belonging to the exponential family generally take the following form,$$f(x;\rho)=h(x)\,\exp({\rho\T Z(x)-A(\rho)})$$ where, $\rho$ is the vector of natural parameters, $Z$ is the vector of sufficient statistics and $A$ is the log-partition function. For a multivariate Gaussian we can obtain each of these as follows:
\begin{align*}
    P(x;\mu,\,\Sigma)&= \dfrac{1}{\sqrt{(2\pi)^n|\Sigma|}}\exp{\bigg(-\dfrac{1}{2}(x-\mu)\T \Sigma^{-1}(x-\mu)\bigg)}\\
    &= \underbrace{\dfrac{1}{\sqrt{(2\pi)^n|\Sigma|}}\exp{\bigg(-\dfrac{1}{2}x\T \Sigma^{-1}x\bigg)}}_{h(x)}\exp{\bigg(x\T \Sigma^{-1}\mu -\dfrac{1}{2}\mu\T \Sigma^{-1}\mu\bigg)}
    =h(x)\exp{(\rho\T Z(x) - A(\rho))}
\end{align*}where, $\rho=\Sigma^{-1/2}\mu,\, Z = \Sigma^{-1/2}x$ and $A(\rho)=\dfrac{1}{2}\mu\T\Sigma^{-1}\mu$. 
Before we compute the gradient we observe the following regarding the log-partition function $A$, 
$$P(x;\mu,\,\Sigma)=h(x) \exp{\big(\rho\T \vZ(x)\big)}\cdot \exp{\big(-A(\rho)\big)} = \dfrac{h(x) \exp{\big(\rho\T \vZ(x)\big)}}{\exp{\big(A(\rho)\big)}}$$
For this to be a valid probability distribution, we must have,\begin{align*}
    \exp{\big(A(\rho)\big)} &= \int h(x) \exp{\big(\rho\T \vZ(x)\big)}\, \rd x \\ \implies A(\rho) &= \ln \int h(x) \exp{\big(\rho\T \vZ(x)\big)} \,\rd x \quad (\text{hence the name \textit{log-partition} function})
\end{align*} We can verify that the expression for $A(\rho)$ obtained above for the Gaussian distribution agrees with this definition of the log-partition function. 
\begin{align*}
    A(\rho) &= \ln \int h(x) \exp(\rho\T Z)\, \rd x\\
    &= \ln \int \dfrac{1}{\sqrt{(2\pi)^n|\Sigma|}}\exp{\bigg(-\dfrac{1}{2}x\T \Sigma^{-1}x\bigg)} \exp\big(x\T\Sigma^{-1}\mu\big) \,\rd x\\
    &=\ln \int \underbrace{\dfrac{1}{\sqrt{(2\pi)^n|\Sigma|}}\exp{\bigg(-\dfrac{1}{2}x\T \Sigma^{-1}x + x\T\Sigma^{-1}\mu - \dfrac{1}{2}\mu\T\Sigma^{-1}\mu\bigg)}}_{f(x;\rho)} \exp{\bigg(\dfrac{1}{2}\mu\T\Sigma^{-1}\mu\bigg)} \,\rd x\\
    &=\ln \int f(x;\rho)\exp{\bigg(\dfrac{1}{2}\mu\T\Sigma^{-1}\mu\bigg)} \,\rd x = \ln \exp{\bigg(\dfrac{1}{2}\mu\T\Sigma^{-1}\mu\bigg)} \int f(x;\rho) \, \rd x = \dfrac{1}{2}\mu\T\Sigma^{-1}\mu
\end{align*}

\par Now it is common practice to simply optimize the mean $\mu$ alone and update the variance using an empirical estimate, which is what we adopt in our algorithm as well. In that case, we are interested in the gradient with respect to $\mu$ alone.  Returning back to the derivation of the gradient update and considering the $f(x;\rho)$ to be the Gaussian distribution, we have the following derivation for the gradient, 
\begin{align*}
    \nabla_\rho \ln f(x;\rho) &= \nabla_\rho(\ln h(x) + \rho\T Z- A(\rho))\\
    &= Z - \nabla_\rho A(\rho) \\
    &= \Sigma^{-1/2}x - \dfrac{1}{2} \nabla_\rho\big\{ (\Sigma^{-1/2}\mu)\T(\Sigma^{-1/2} \mu) \big\}\\
    &= \Sigma^{-1/2}x - \Sigma^{-1/2} \mu \quad (\text{because } \rho = \Sigma^{-1/2}\mu) \\
    &= \Sigma^{-1/2}(x-\mu)
\end{align*}
Substituting this back into the expression for the gradient of the objective we get the following gradient ascent update for the parameter $\rho$, 
\begin{align*}
    &\rho^{k+1} = \rho^k + \alpha \dfrac{\Eb\big[S\big(F(x)\big)\big(\Sigma^{-1/2}(x-\mu)\big)\big]}{\Eb\big[S\big(F(x)\big)\big]}\\
    &\text{Using } \rho=\Sigma^{-1/2}\mu, \text{ we have, }  \Sigma^{-1/2} \mu^{k+1} = \Sigma^{-1/2} \mu^k + \alpha \Sigma^{-1/2} \dfrac{\Eb\big[S\big(F(x)\big)(x-\mu)\big]}{\Eb\big[S\big(F(x)\big)\big]}\\
    &\text{Therefore, } \mu^{k+1} = \mu^k + \alpha\dfrac{\Eb\big[S\big(F(x)\big)(x-\mu)\big]}{\Eb\big[S\big(F(x)\big)\big]}
\end{align*}
\subsection{NOVAS algorithm}
\label{sec:novas_algo}

In this section, we show the algorithmic implementation of the NOVAS module. Since the goal of this module is to solve the problem proposed in eqn. \eqref{eqn:hamiltonian_minimization}, we must provide all values and functions needed for the computation of the Hamiltonian such as the current state vector, the gradient of the Value function at the current time step and the system's drift vector. In addition to these quantities, we also have tunable hyperparameters, referred to in algorithm \ref{alg:first_exit} as $NOVAS\_inputs$, that directly affect the performance of the algorithm. These values include the initial sampling mean and variance $(\mu_0, \Sigma)$, a scalar learning rate $(\alpha)$, user defined number of NOVAS samples and NOVAS inner-loop iterations $(M, N)$, some arbitrarily small positive number $(\varepsilon)$ indicating minimum variance, and a user-defined shape function $(S)$. The quantity $\varepsilon$ and function $S$ are set to improve stability of the algorithm while all other values directly affect convergence speed and accuracy of the control solution. The exact values used to obtain the presented simulation results are presented in table \ref{table:hyperparams}.

\begin{algorithm}[h!]
\caption{NOVAS\_LAYER}
\label{alg:novas_layer}
\begin{algorithmic}[1]
\Function{NOVAS\_LAYER}{$\vx[b,t],\, V_\vx[b,t],\, \mathcal{H},\,f$, NOVAS\_inputs:, initial sampling mean and variance ($\mu_0, \Sigma$), learning rate ($\alpha$), shape function ($S$), number of samples ($M$), number of iterations ($N$), small positive number ($\epsilon$)}

\Statex
\State \textbf{Initialize}: $\mu \leftarrow \mu_0$

\Statex \textit{(Obtain an optimal control policy by minimizing the Hamiltonian)}
\For{$n=1:N-1$ \textbf{\textit{(off-graph operations)}}}
\State $(\mu, \, \Sigma) \leftarrow\,$\textbf{NOVAS\_STEP}$\,(\vx[b,t],\, V_\vx[b,t],\, \mathcal{H}, \, f, \, \mu, \, \Sigma, \, \alpha, \, S, \, M, \, \epsilon)$ \Comment{Algorithm \ref{alg:novas_step}}

\EndFor

\State $(\mu, \, \Sigma) \leftarrow\,$\textbf{NOVAS\_STEP}$\,(\vx[b,t],\, V_\vx[b,t],\, \mathcal{H}, \, f, \, \mu, \, \Sigma, \, \alpha, \, S, \, M, \, \epsilon)$

\State $\vT^* \leftarrow \bigg(\mu_1, \mu_2, \sqrt{(\mu_3)^2 -(\mu_1)^2 - (\mu_2)^2} \bigg)$

\State \Return $(\vT^*)$
\EndFunction
\end{algorithmic}
\end{algorithm}

During each NOVAS iteration, we approximate the gradient through sampling. To do this, we sample $M$ different values of horizontal thrust and thrust norm using univariate Gaussian distributions by using a vector of mean values $\mu$ and a covariance matrix $\Sigma$ that is a diagonal matrix. During initialization, the mean vector can be populated using random values within the admissible control set. However, in our case, we have set such values to be at the lower bound of the valid thrust levels with zero lateral thrust (i.e.: $\mu = (0,0,\rho_1)$) for the first ($ k=0$) time step, and use the optimal control from previous time step (i.e., $\mu_k^* = \mu_{k-1}^*$) for all subsequent time steps $ k >0$. Note that the first $N-1$ iterations of NOVAS are \textit{off-graph operations}, meaning that they are not part of the deep learning framework's compute graph and therefore not considered during backpropagation. A compute graph is built to approximate gradients, by means of automatic differentiation, of the loss function with respect to the weights of the neural network. Taking the first $N-1$ iterations \textit{off-the-graph} can be done to warm-start the last iteration which is performed \textit{on-the-graph}. This procedure has negligible effect on the training of the neural network and can be performed because NOVAS does not overfit to the specific number of inner-loop iterations as demonstrated in \cite{exarchos2020novas}. By performing the first $N-1$ operations of NOVAS \textit{off-the-graph} we significantly reduce the size of the compute graph speeding up training and enabling us to use this approach to train policies for long time horizons.

\begin{algorithm}[h!]
\caption{NOVAS\_STEP}
\label{alg:novas_step}
\begin{algorithmic}[1]
\Function{NOVAS\_STEP}{$\,\vx[b,t],\, V_\vx[b,t],\, \mathcal{H}, \, f, \, \mu, \, \Sigma, \, \alpha, \, S, \, M, \, \epsilon$} 

\State \textit{Generate} $M$ \textit{control samples:} $(\bar{x}^m, \, \delta \bar{x}^m)  \leftarrow \textbf{SAMPLE}(\mu, \Sigma),  \quad  m=1,...,M $ \Comment{Algorithm \ref{alg:sample_controls}}

\State \textit{Transform:} $\vT^m \leftarrow \bigg(\bar{x}_1^m, \bar{x}_2^m, \sqrt{(\bar{x}_3^m)^2 -(\bar{x}_1^m)^2 - (\bar{x}_2^m)^2} \bigg)$

\For{$m=1:M$ \textbf{\textit{(vectorized operations)}}}
\State \textit{Evaluate:} $F^m = -\mathcal{H}(\,\vx[b,t],\, V_x [b,t], \, \vT^m, f)$
\Comment{using eqn. \eqref{eqn:simplified_hamiltonian}}
\State \textit{Shift:} $F^m = F^m - \min_m(F^m)$
\State \textit{Apply shape function:} $S^m = S(F^m)$
\State \textit{Normalize:} $S^m = S^m/ \sum^{M}_{m=1} S^m$
\EndFor

\Statex \textit{(Perform control mean and variance update)}

\State $\mu = \mu + \alpha \sum^{M}_{m=1} S^m \delta \bar{x}^m$
\State $\delta \bar{x}^m \leftarrow \bar{x}^m - \mu$
\State $\Sigma = diag \bigg( \sqrt{\sum^{M}_{m=1} S^m (\delta \bar{x}^m)^2 + \epsilon} \bigg)$

\State \Return ($\mu, \Sigma$)
\EndFunction
\end{algorithmic}
\end{algorithm}

% Add constrained controls sampling algo here
\begin{algorithm}[h!]

\caption{Sampling with control constraints for NOVAS}
\label{alg:sample_controls}

\begin{algorithmic}[1]

\Function{SAMPLE}{$\mu, \Sigma$} 
\State \textbf{Given}: $\rho_1, \, \rho_2, \,$ and $\theta$
\State \textit{Compute}: $\rho_3 \leftarrow \sqrt{\frac{\rho_1^2}{2 \cdot \sin^2 \theta}}$

\State Sample: $x \sim \mathcal{N}(\mu,\, \Sigma)$
\State Project samples: 
\Statex $\quad\quad\quad\quad\bar{x}_1 = \text{Proj}_{[-\rho_1/2,\, \rho_1/2]}(x_1)$ 
\Statex $\quad\quad\quad\quad\bar{x}_2 = \text{Proj}_{[-\rho_1/2,\, \rho_1/2]}(x_2)$
\Statex $\quad\quad\quad\quad\bar{x_3} = \text{Proj}_{[\max(\rho_1,\, \rho_3), \, \rho_2]}(x_3)$
% \State Compute samples of $\vT_3 = \sqrt{||\bar{\mathbf{\vT}}||^2 - \bar{\vT}_1^2 - \bar{\vT}_2^2}$ \Comment{By limiting $\vT_1$ and $\vT_2$ this is always positive}

\State $\bar{x} \leftarrow (\bar{x}_1, \, \bar{x}_2, \, \bar{x}_3)$
\State $\delta \bar{x} \leftarrow \bar{x} - \mu$
\State \Return ($\bar{x}, \delta \bar{x}$)

\EndFunction
\end{algorithmic}
\end{algorithm}

\subsection{Simulation hyperparameters and compute resources}
\label{sec:sim_hyperparam} Our simulations were coded in PyTorch \cite{paszke2019pytorch} and run on a desktop computer with an Intel Xeon E5-1607 V3 3.1GHz 4-core CPU and a NIVIDIA Quadro K5200 Graphics card with 8GB VRAM. 
In table \ref{table:hyperparams} below, we list values of some of the other hyperparameters not mentioned in the main body of this paper.
\begin{table}[h!]
\begin{center}
    \begin{tabular}{||c|c||}
    \hline
        \textbf{Hyperparameter name} & \textbf{Hyperparameter value} \\
        \hline\hline
         $(\rho_1,\,\rho_2)$ & $(4.97\times 10^3,\,1.334\times 10^4)$\\
         (dry-mass, initial mass)=($m_d,\,m_0$) & (1700 kg, 1905 kg)\\
         Minimum admissible glideslope angle, $\gamma$ & $\dfrac{\pi}{4}$\\
         Glide-slope cost coefficients, $(q_+,\, q_-)$ & $(1.0,\,0.005)$\\
         Acceleration due to gravity, g & 3.7144 $m/s^2$\\
         Fuel-consumption rate, $\alpha$ & $4.85\times 10^{-4}$\\
         Tolerance for landing/crash, $h_\text{tol}$ & $10^{-3}$ m\\
         Terminal z-velocity cost coefficients ($c_{v_z +},\,c_{v_z -}$) & (10.0, 1.0)\\
         Diffusion matrix for dynamics, $\Sigma$ & $10^{-4}\cdot \vI_{3\times 3}$\\
         Initial altitude & 80 m\\
         Initial vertical velocity & -10 m/s \\
         Radius of base of glide-slope cone, $rad$ & 80 m\\
         Initial horizontal velocity, ($r_1(0),\,r_2(0)$) & (0, 0) m/s\\
         Number of LSTM layers & 2\\
         Hidden and cell state neurons per layer & 16\\
         Optimizer & Adam \\
         NOVAS shape function, $S(\cdot)$ & $\exp(\cdot)$\\
         NOVAS initial sampling variance, $\Sigma$ & \textbf{diag}($500^2,\,500^2,\,1000^2$)\\
         NOVAS initial sampling mean & $[0.0,\,0.0,\,5000]$\\  
         NOVAS iteration learning rate, $\alpha$ & 1.0 \\ maximum allowable angle
between the $\vT$ and $\hat{\vn}$, $\theta$ & $\dfrac{\pi}{4}$\\ 
         \hline
    \end{tabular}
    \caption{Hyperparameter values}
    \label{table:hyperparams}
\end{center}
\end{table}

\bibliography{sample}
\end{document}